\documentstyle[12pt,epsfig]{article}
\def\beq{\begin{equation}}
\def\eeq{\end{equation}}
\def\bea{\begin{eqnarray}}
\def\eea{\end{eqnarray}}
\def\bq{\begin{quote}}
\def\eq{\end{quote}}

\newcommand{\rr}{\mbox{\boldmath $r$}}
\newcommand{\rrn}{\mbox{$r$}}

\newcommand{\lsim}{\raisebox{-0.5mm}{$\stackrel{<}{\scriptstyle{\sim}}$}}

\def\gappeq{\mathrel{\rlap {\raise.5ex\hbox{$>$}}
{\lower.5ex\hbox{$\sim$}}}}

\def\lappeq{\mathrel{\rlap{\raise.5ex\hbox{$<$}}
{\lower.5ex\hbox{$\sim$}}}}

\def\Toprel#1\over#2{\mathrel{\mathop{#2}\limits^{#1}}}

\begin{document}
\pagestyle{empty}
\begin{flushright}
%\verb!version 1.8!
%{CERN-TH/2001-265}\\
DESY 03-016\\
hep-ph/0302079\\  
\end{flushright}
%\vspace*{5mm}
\begin{center}
{\bf Deeply virtual Compton scattering and saturation approach} 
\\ 
\vspace*{1cm}
L. Favart $^{\star}$, M.V.T. Machado $^{\star\star}$\\
\vspace{0.3cm}
{$^{\star}$ \rm IIHE - CP 230, Universit\'e Libre de Bruxelles\\
  1050 Brussels, Belgium}\\

$^{\star\star}$ \rm High Energy Physics Phenomenology Group, GFPAE, IF-UFRGS\\
 Caixa Postal 15051, CEP 91501-970, Porto Alegre, RS, Brazil\\
 and \\
 \rm Instituto de F\'{\i}sica e Matem\'atica, Universidade 
 Federal de Pelotas\\
 Caixa Postal 354, CEP 96010-090, Pelotas, RS, Brazil

\vspace*{1cm}
{\bf ABSTRACT} 
\end{center}

\vspace*{1mm} \noindent We investigate the deeply virtual Compton
scattering (DVCS) in the color dipole approach, implementing the dipole
cross section through the saturation model, 
%cross section through the saturation model~\cite{Golec-Biernat:1998js}, 
which interpolates
%1.6 cross section through the saturation model, which interpolates
successfully between soft and hard regimes. The imaginary and real part of
the DVCS amplitude are obtained and the results are compared to the
available data.
%1.6  in a parameter-free way.

\vspace*{1cm}
\noindent
\rule[.1in]{16.5cm}{.002in}

\vspace{-2cm}
\setcounter{page}{1}
\pagestyle{plain}

\vspace{1cm}

\section{Introduction}

% Initial Considerations % ------ 
The investigation of hard exclusive
reactions in the Bjorken limit is a reliable acces of obtaining information into
details on the structure of the nucleon which could not be obtained
considering inclusive deep inelastic processes.
The probe provided by the photon works as a clean tool reactions
in order to extract reliable knowledge on the substructure of strongly
interacting particles complementar to exclusive vector meson
production~\cite{Caldwell:2001ky}.
The recent data from the DESY $ep$ collider HERA on
exclusive diffractive virtual Compton process \cite{Adloff:2001cn,ZEUS02} (DVCS)
at large $Q^2$ becomes an important source to study the partons, in
particular gluons, 
inside the proton for nonforward kinematics and its relation with
the forward one. These features are encoded in the formalism of the
so-called skewed parton distributions, where dynamical correlations
between partons with different momenta are taken into account. A
considerable interest of the DVCS process lies in the particular
access it gives to these generalized parton distributions (GDP) through
the interference term with the Bethe-Heitler process. The high energy
situation at HERA gives the important opportunity to constrain the gluon
contribution to GPDs and to study the evolution in virtuality $Q^2$ of the
quark and gluon distributions. On the other hand, recently the color
dipole formalism has provided a simultaneous description of photon
induced process. The inclusive deep inelastic reaction and the photon
diffractive dissociation has been successfully described and the study of
other exclusive processes such as DVCS is an important test of the color dipole
approach.

% DVCS LO and NLO figures
% ------

\begin{figure}[t]
\begin{tabular}{cc}
\psfig{file=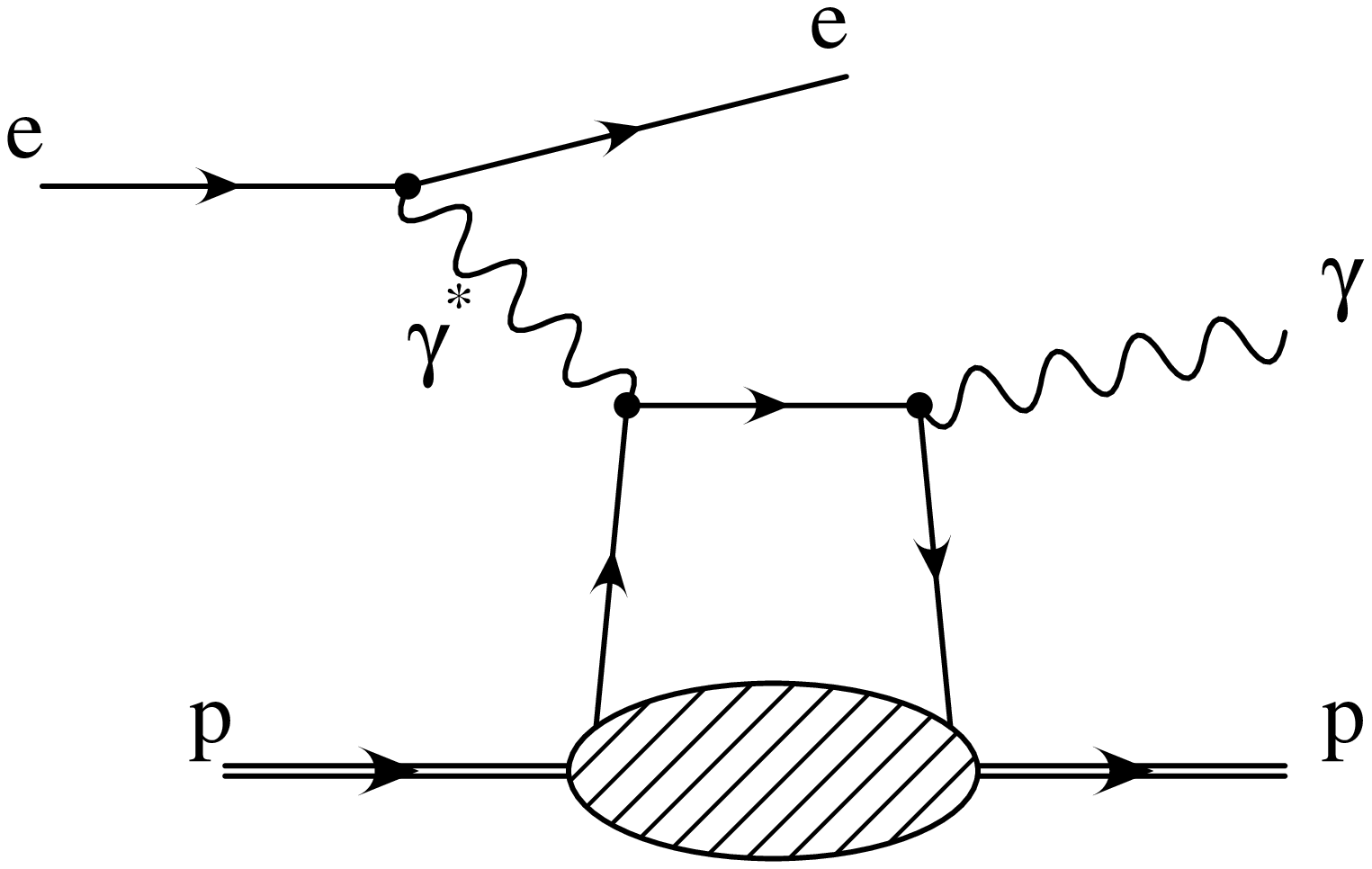,width=70mm} & \psfig{file=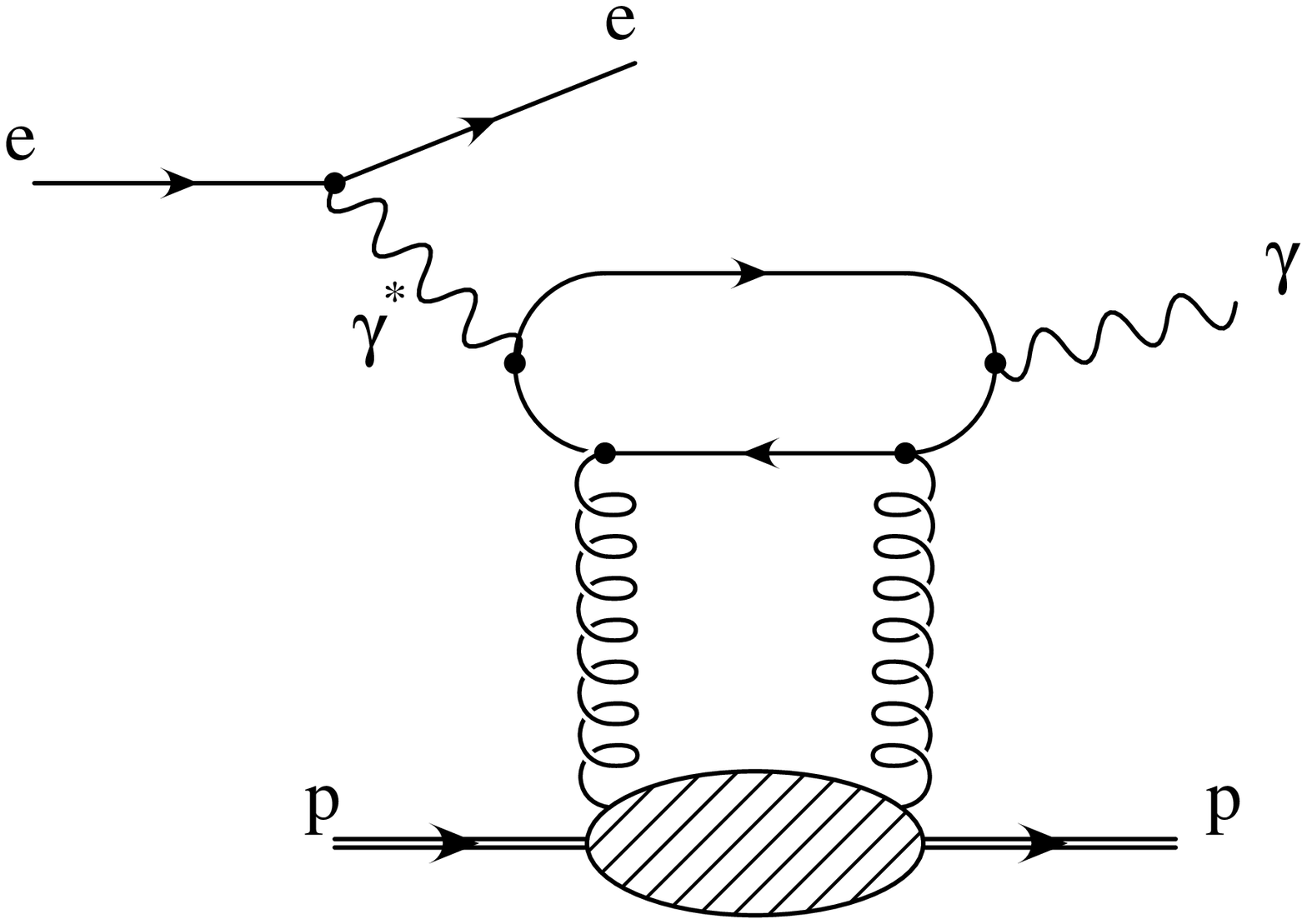,width=70mm}\\
\end{tabular}
 \caption{The two leading DVCS QCD diagrams. The LO diagram in the 
 QCD picture on the left graph and the  NLO diagram on the right.}
 \label{fig:diag1}
\end{figure}

% Dipole
% ------
The color dipole model of diffraction is a simple unified picture of
diffractive processes which provides a large phenomenology on the 
Deep Inelastic Scattering (DIS) regime.
In the proton rest frame, the DVCS process can be seen as 
a succession in time of three factorisable subprocesses: i) 
the photon fluctuates in a quark-antiquark pair, ii) this 
coulour dipole interacts with the proton target, iii) the quark pair
annihilates in a real photon.
Using as kinematic variables the $\gamma^* p$ c.m.s. energy squared $s=W^2=(p+q)^2$, 
where $p$ and $q$ are the proton and the photon
momenta respectively, the photon virtuality squared $Q^2=-q^2$ and
the Bjorken scale $x=Q^2/(W^2+Q^2)$, 
the corresponding amplitude at zero momentum transfer reads as, 
\begin{eqnarray} {\cal I}m
\, {\cal A}_{T,L} (s,Q^2,\,t=0) & = & \sigma^{\gamma^* p}_{T,\,L}(x,\,Q^2)= \int_0^1
dz\, \int d^2\rr \, |\Psi_{T,\,L} (z,\,\rr,\,Q^2)|^2 \, \sigma_{dip}
(\tilde{x},\,\rr^2)\,,\\
 |\Psi_{T} (z,\rr,\,Q^2)|^2 & = &  \frac{6\alpha_{\mathrm{em}}}{4\,\pi^2} \, 
 \sum_f e_f^2 \, \left\{[z^2 + (1-z)^2]\, \varepsilon^2 \,K_1^2(\varepsilon \,\rrn) 
 + m_f^2 \, \,K_0^2(\varepsilon\, \rrn)   
 \right\}\,,\label{wtrans}\\
 |\Psi_{L} (z,\rr,\,Q^2)|^2 & = & \frac{6\alpha_{\mathrm{em}}}{\pi^2} \,
\sum_f e_f^2 \, \left\{Q^2 \,z^2 (1-z)^2 \,K_0^2(\varepsilon\, \rrn)  
\right\}\,, \end{eqnarray} 
where $\Psi_{T,\,L}$ are the photon wave
function for transverse ($T$) and longitudinal ($L$) photons,
respectively. They are well known from usual QED coupling, where $\rr$
defines the relative transverse separation of the pair (dipole) and $z$,
$(1-z)$, are the longitudinal momentum fractions of the quark (antiquark).
The auxiliary variable $\varepsilon^2=z(1-z)\,Q^2 + m^2_f$ depends on the
quark mass, $m_f$. The $K_{0,1}$ are the McDonald functions and summation
is performed over the quark flavors.

% Sigma_dipole
% ------------
The dipole cross section $\sigma_{dip}$ describing the dipole-proton
interaction is substantially affected by non-perturbative content.
There are several phenomenological implementations for this quantity. The
main feature in a successful model is to be able to match the soft (low
$Q^2$) and hard (large $Q^2$) regimes in a unified way. This has been
achieved in models using Regge approach \cite{Forshaw:1999uf}, QCD color
transparency phenomena \cite{McDermott:1999fa} and/or hybrid ones mixing
non-perturbative functional approach and Regge phenomenology
\cite{Donnachie:2000px}. Phenomenological analysis including a parton
saturation phenomenon at high energies has been considered, for instance
in \cite{GayDucati:2001ey,Betemps:2001he}.
\\

% Saturation
% ----------
In the present work we follow
the quite successful saturation model \cite{Golec-Biernat:1998js}, which
interpolates between the small and large dipole configurations, providing
color transparency behavior, $\sigma_{dip}\sim \rr^2$, as $\rr \rightarrow
0$ and constant behavior, $\sigma_{dip}\sim \sigma_0$, at large
%1.6 0$ and confinement behavior, $\sigma_{dip}\sim \sigma_0$, at large
dipoles. The parameters of the model have been obtained from an adjustment to
small $x$ HERA data. Its free-parameter application to diffractive DIS
has been also quite successful \cite{Golec-Biernat:1999qd}.   The
parametrisation for the dipole cross section takes the eikonal-like form,
\begin{eqnarray} \sigma_{dip} (\tilde{x}, \,\rr^2) & = & \sigma_0 \,
\left[\, 1- \exp \left(-\frac{\,Q_s^2(x)\,\rr^2}{4} \right) \, \right]\,,
\label{gbwdip}\\ Q_s^2(x) & = & \left( \frac{x_0}{\tilde{x}}
\right)^{\lambda} \,\,\mathrm{GeV}^2\,, 
\end{eqnarray} 
where the
saturation scale $Q_s^2$ defines the onset of the saturation phenomenon,
which depends on energy. The parameters were obtained from a fit to the
HERA data producing $\sigma_0=23.03 \,(29.12)$ mb, $\lambda= 0.288 \,
(0.277)$ and $x_0=3.04 \cdot 10^{-4} \, (0.41 \cdot 10^{-4})$ for a
3-flavor (4-flavor) analysis~\cite{Golec-Biernat:1998js}. 
An additional parameter is the effective
light quark mass, $m_f=0.14$ GeV. It is
%1.6 light quark mass, $m_f=0.14$ GeV, consistent with the pion mass value. It is
worth to mention that the quark mass plays the role of a regulator for the
photoproduction ($Q^2=0$) cross section. The charm quark mass is
considered to be $m_c=1.5$ GeV. A smooth transition to the photoproduction
limit is obtained with a modification of the Bjorken variable as,
\begin{eqnarray} \tilde{x}= x\, \left( \, 1+ \frac{4\,m_f^2}{Q^2}
\,\right)\,. \end{eqnarray}
%1.6 The saturation model is suitable in the region below $x=0.01$ and the
%1.6 large $x$ limit needs still a consistent treatment.  Here, we supplement
%1.6 the dipole cross section, Eq. (\ref{gbwdip}), with a threshold factor
%1.6 $(1-x)^{n_{\mathrm{thres}}}$, taking $n_{\mathrm{thres}}=5$ for a 3-flavor
%1.6 analysis and $n_{\mathrm{thres}}=7$ for a 4-flavor one. This is in order
%1.6 to obtain consistent predictions at low energy and completeness.

% BGK model
% ---------

 Recently, a new implementation of the model including QCD evolution
\cite{Bartels:2002cj} (labeled in the following BGK model) has appeared.
It will be considered in the calculation to be presented in
the last section. Now, the dipole cross section depends on the gluon
distribution in the following way, 
\begin{eqnarray}
 \sigma_{dip} (\tilde{x}, \,\rr^2) & = & \sigma_0 \, \left[\, 1- \exp
\left(-\frac{\,\pi^2\,\rr^2\,\alpha_s(\mu^2)\,\tilde{x}\,G(\tilde{x},\mu^2)}{3\,\sigma_0}
\right) \, \right]\,,\\ x\,G(x,\mu^2) & = &
A_g\,x^{-\lambda_g}\,(1-x)^{5.6}\,, 
\label{bgkdip} 
\end{eqnarray} 
where $\mu^2=C/\rr^2 + \mu_0^2$. The parameters 
($C$ and $\mu_0^2$) are
determined from a fit to DIS data. The flavor number is taken to be equal to 3.
The overall normalization $\sigma_0=23.03$ mb is kept fixed (labeled fit 1 in
Ref. \cite{Bartels:2002cj}). The function $G(x,\mu^2)$ is evolved with the
leading order DGLAP evolution equation for the gluon density with initial
scale $Q_0^2=1$ GeV$^2$. The improvement preserves the main features of the
low-$Q^2$ and transition regions, while providing QCD evolution in the
large-$Q^2$ domain.

In this paper we test the validity of the saturation model 
in the particular case of the DVCS:
$\gamma^* p \rightarrow \gamma p$, where the final photon is real.
The DVCS consists in the simplest exclusive diffractive process as 
it does not imply uncertainties due to the composite meson wave function 
in the final state like in the vector meson production, 
$\gamma^* p \rightarrow V p$. 
On the other hand, as we will see, the restriction to the 
transverse part of the photon wave function (at the leading twist), due to the
real photon, enhances
the contribution of larger dipole configurations and therefore
the sensitivity, w.r.t the inclusive DIS case,
to the soft content of the proton and the transition between the hard and soft regimes.
This is why we believe that the DVCS process provides a particularly relevant
test, in a parameter free way, of the saturation model. 
The recent cross section measurements from
H1~\cite{Adloff:2001cn} and ZEUS~\cite{ZEUS02} at HERA allow us now to confront the saturation 
%1.6 H1~\cite{Adloff:2001cn} and ZEUS at HERA allow us now to confront the saturation 
model prediction.

% paper outlook
% ----------

The paper is organized as follows. In the next section we address the main formulae of
the DVCS process in the color dipole picture applying the dipole cross section given
by the saturation model in order to investigate and to calculate the imaginary and
real part of the referred amplitude. In the last section we compare the results
obtained with the DVCS measurement of the H1 Collaboration, taking into
account the three(four)-flavor analysis model. The results obtained using the recent
QCD improvement to the original model are also presented. There, we discuss the
different characteristics coming from the distinct analysis and present a qualitative
study of the DVCS cross section using the saturation model.

%In the well known DIS process, both
%initial and produced photon has equal virtuality and the subjacent
%phenomenology allows to constrain the correspondent parameters in an
%underlying approach describing it. The DVCS process, on the other hand, is
%related with the vector meson production, $\gamma^* p \rightarrow V p$,
%where now one has the production of a real photon, avoiding further
%knowledge about the meson wave function which is the main source of
%uncertainty in those reactions.

\section{Deeply virtual Compton scattering in the color dipole picture}

In this section we establish the deeply virtual Compton scattering amplitude 
and the profile function expressions.

Applying the color dipole approach to the DVCS process, 
the imaginary part of the amplitude at zero momentum transfer reads as, 
\begin{eqnarray} 
 & & {\cal I}m\, {\cal A}\,(s,Q^2,t=0)  = \int_0^1 dz\, \int d^2\rr \,
\Phi_T(z,\,\rr,\,Q^2)\,\sigma_{dip}(\tilde{x},\,\rr^2)\,,\label{dvcsdip}\\
& & \Phi_T (z,\,\rr,\,Q^2)  \equiv \Psi_T^*(z,\,\rr,\,Q_1^2=Q^2)\,
\Psi_T(z,\,\rr,\,Q_2^2=0)\,,\\ & & \Phi_T =
\frac{6\alpha_{\mathrm{em}}}{4\,\pi^2} \, \sum_f e_f^2 \, \left\{[z^2 +
(1-z)^2]\, \varepsilon_1 \,K_1 (\varepsilon_1 \,\rrn) \,\varepsilon_2
\,K_1 (\varepsilon_2 \,\rrn) + m_f^2 \, \,K_0(\varepsilon_1\,
\rrn)\,K_0(\varepsilon_2\, \rrn)  \right\}\,,\label{wdvcstrans}
\end{eqnarray} 
where $\varepsilon^2_{1,\,2}= z(1-z)\,Q_{1,\,2}^2 + m_f^2$.
\\

The relative contributions from dipoles of different sizes can be analyzed
with the weight (profile)  function, \begin{eqnarray} W(\rr,Q^2) =
\int_0^1 \,dz\; r \; \Phi_T(z,\rr,Q^2) \; \sigma_{dip}\,(\tilde{x},\,\rr^2)\,.
\end{eqnarray}

%The initial
%photon has virtuality $Q_1^2=Q^2$ and the final state photon has
%virtuality equal zero, $Q_2^2=0$. 

\begin{figure}[t]
\begin{tabular}{cc}
\psfig{file=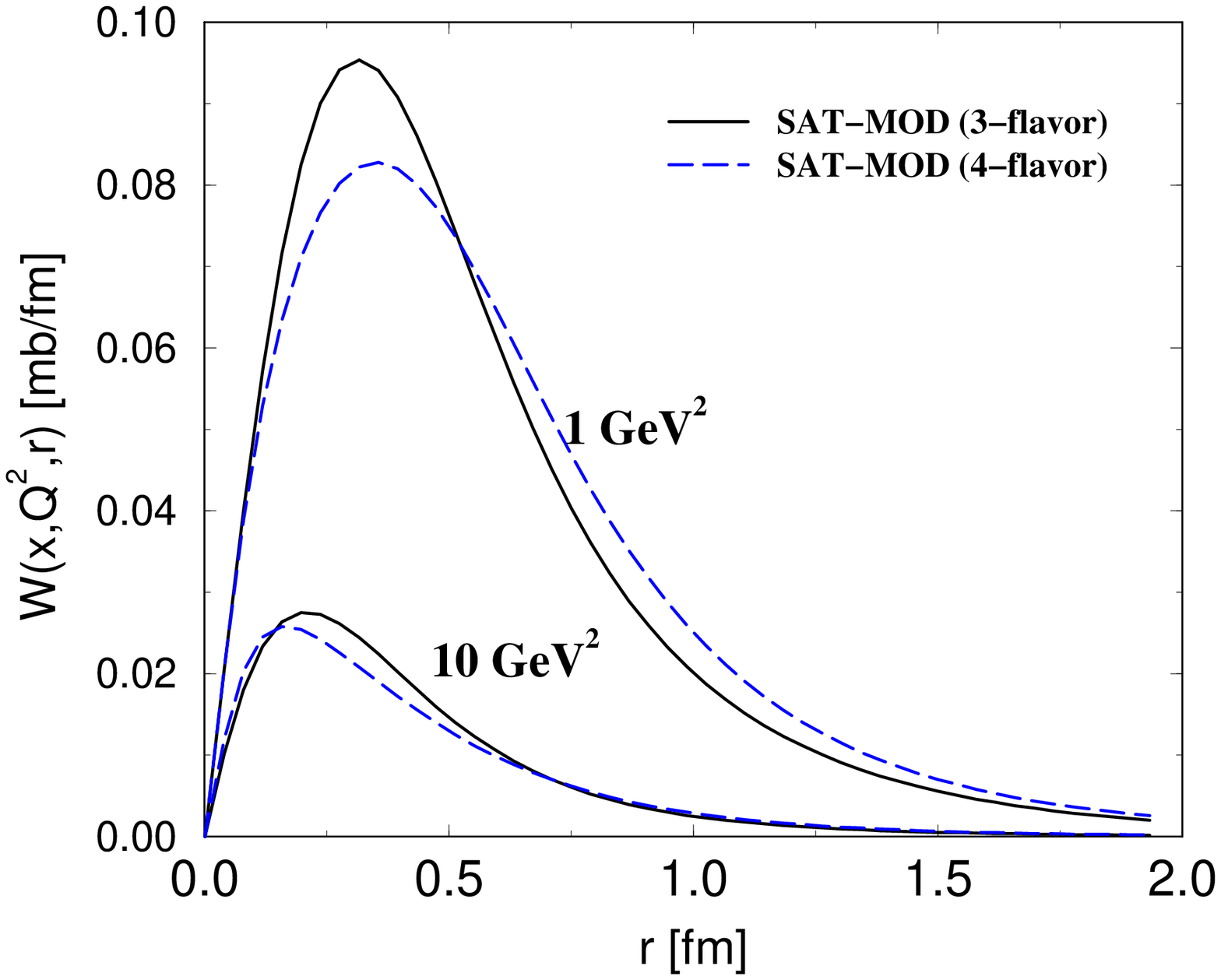,width=80mm} & \psfig{file=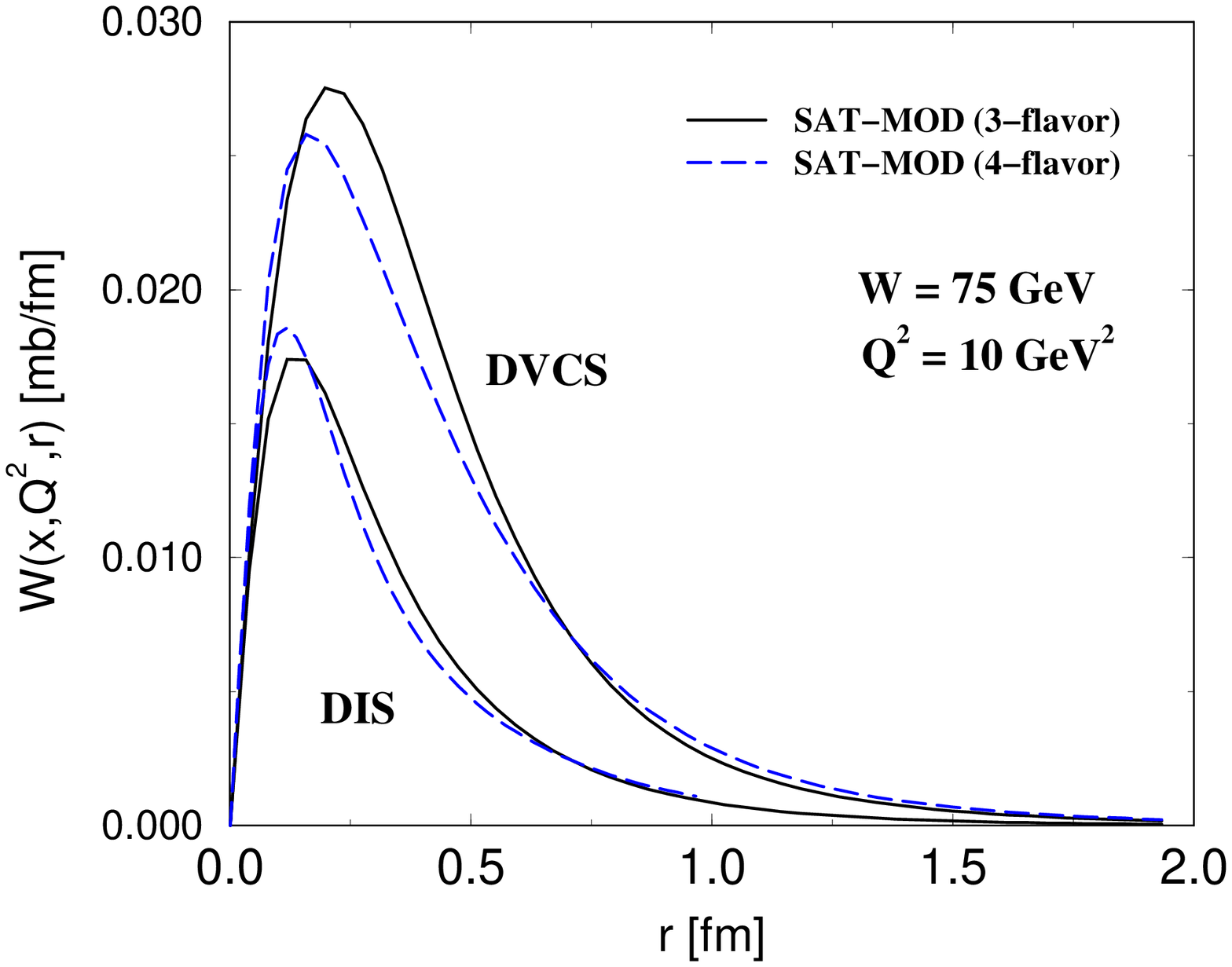,width=80mm}\\
\end{tabular}
\caption{\it {\bf (a)} The weight function $W(\rr,Q^2)$ as a function of the 
 dipole size $\rr$ at fixed c.m.s energy $W=75$ GeV for two virtualities: 
 $Q^2=1$ GeV$^2$ and $Q^2=10$ GeV$^2$. The solid lines correspond to a 3-flavor 
 analysis whereas the long-dashed ones are for the 4-flavor case. {\bf (b)} 
 Comparison between the profile function for DVCS and inclusive DIS processes 
 at $Q^2=10$ GeV$^2$, using the same notation.}
\label{rprofile} 
\end{figure} 

In Fig.~(\ref{rprofile}-a), the weight function $W(\rr,Q^2)$ is shown
as a function of the dipole size $r$ at fixed c.m.s energy $W=75$ GeV and
at two virtualities. The dipole cross section is taken from Eq. (\ref{gbwdip}).  
We analyze the $r$-profile for a low virtuality
$Q^2=1$ GeV$^2$ and for a large one, $Q^2=10$ GeV$^2$. Our conclusions are
similar to those in Ref. \cite{McDermott:2001pt}, i.e. as the virtualities
increase the contribution of large dipole configurations diminishes in a
sizable way. The calculation was performed using both a 3-flavor and a
4-flavor analysis. The inclusion of the charm content gives a slightly lower
normalization for the profile and by consequence for the total cross
section. It is worth to mention that the situation is different in the inclusive DIS case,
where the transverse wave function is symmetric on $\varepsilon_{1,2}$
[for instance, see Eqs. (\ref{wtrans}) and (\ref{wdvcstrans})] and the
results including charm slightly differ from those using only light
quarks. This is shown in Fig.~(\ref{rprofile}-b) where one
compares the DVCS and inclusive DIS profiles at $Q^2=10$ GeV$^2$ using light quarks
and adding charm contribution. The impact of charm is smaller in the
inclusive DIS case than in the DVCS process. 
Furthermore, it is verified that the DVCS
profile selects larger dipole sizes in contrast to the inclusive DIS profile, even
in the analysis including charm. Therefore, DVCS is more sensitive to the
non-perturbative (soft) content of the scattering process.
\begin{figure}[t]
\begin{tabular}{cc}
\psfig{file=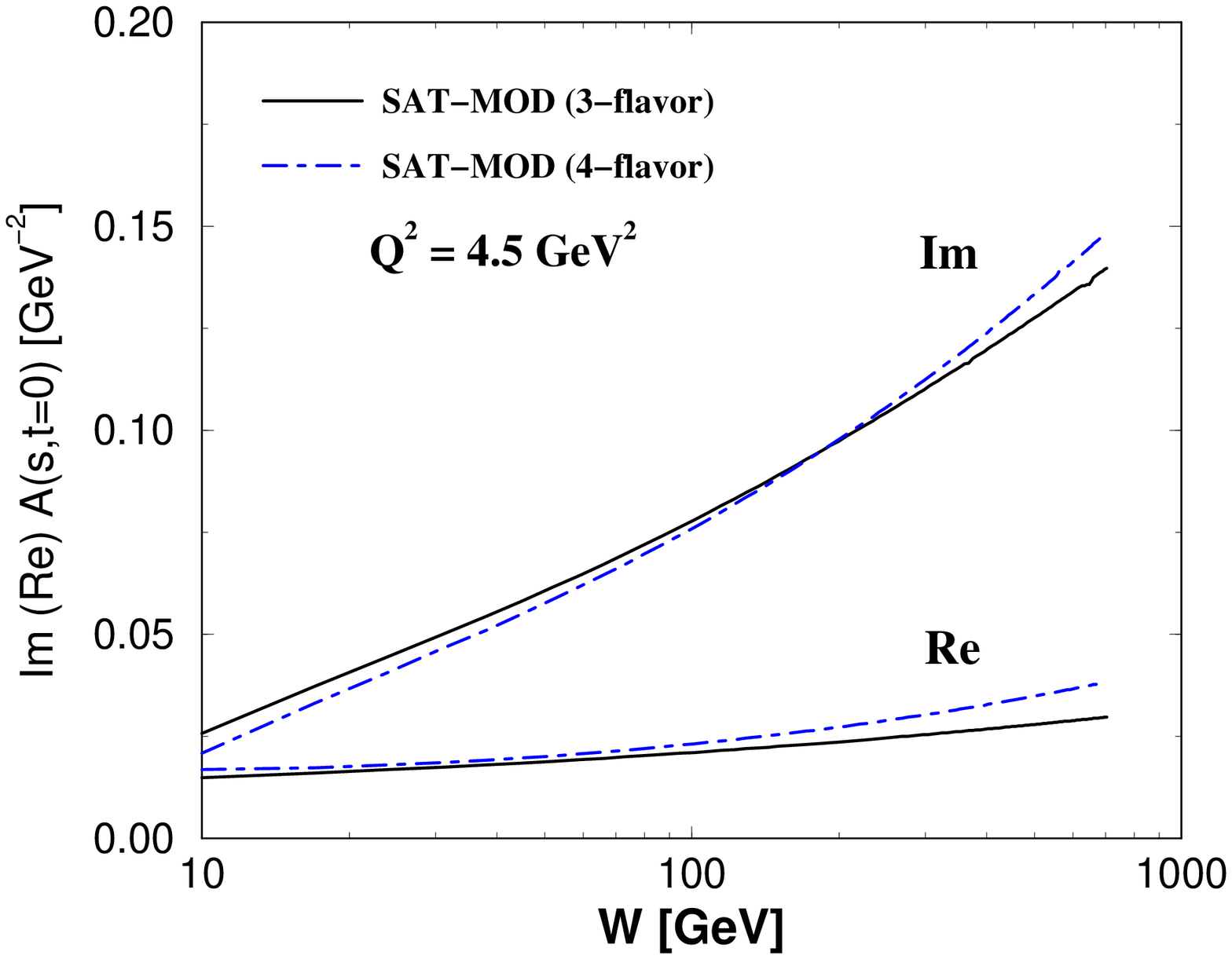,width=80mm} & \psfig{file=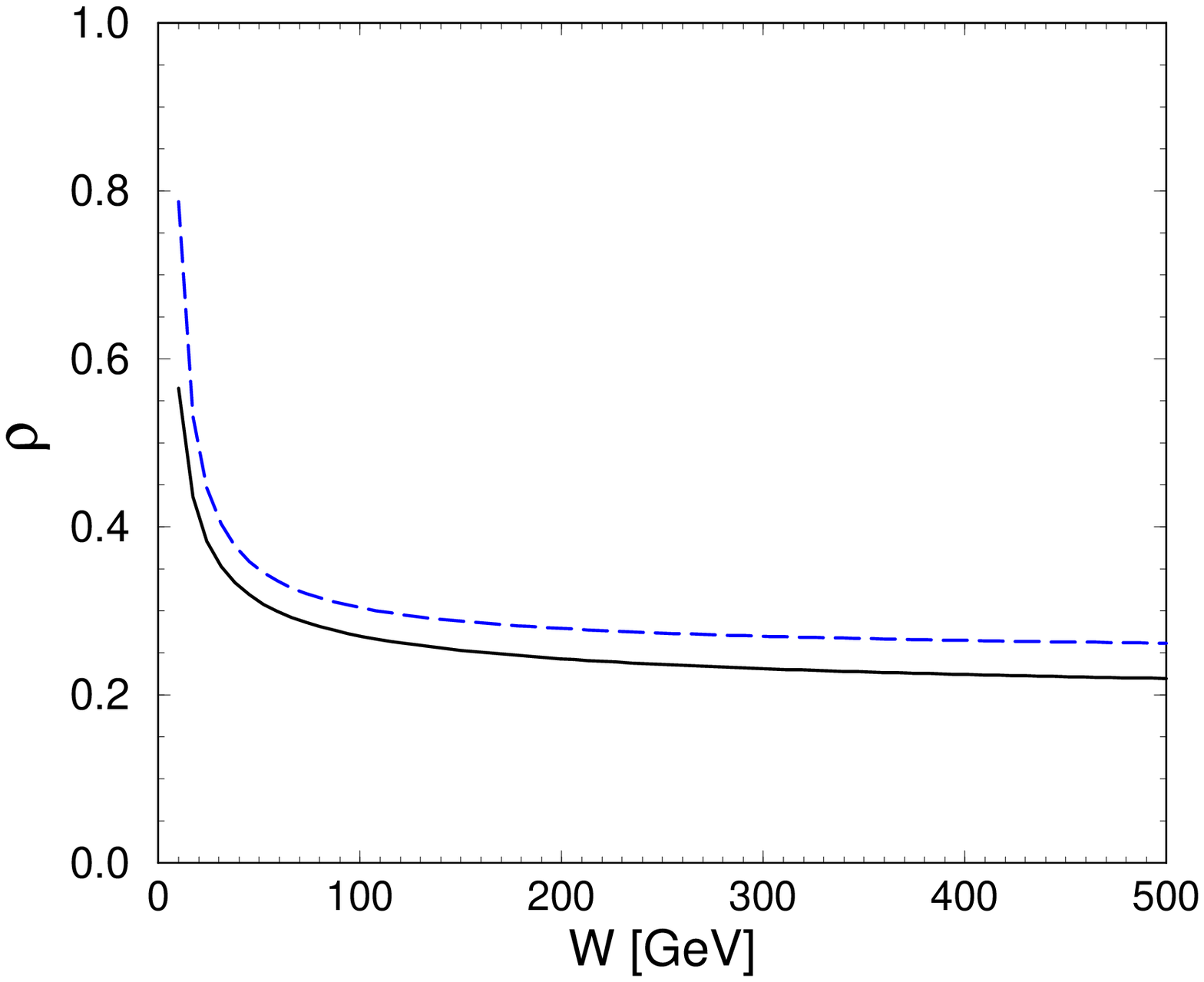,width=80mm}\\
\end{tabular}
\caption{\it {\bf (a)} The imaginary and real part of the DVCS 
 amplitude as a function of the c.m.s. energy $W$ at $Q^2=4.5$ 
 GeV$^2$. {\bf (b)} The ratio real to imaginary part, the 
 parameter $\rho$ as a function of $W$ at  $Q^2=4.5$ GeV$^2$. 
 The solid lines correspond to the 3-flavor analysis and 
 the long-dashed ones to the 4-flavor analysis.}
\label{ampsplot} 
\end{figure} 

In order to compute the real part of the DVCS amplitude in the approach
above, we have considered the dispersion relations as done in of Ref.
\cite{McDermott:2001pt}. In Fig.~(\ref{ampsplot}-a) both the
imaginary and real part of the amplitude using the saturation model 
are shown as a function of c.m.s. energy $W$ at $Q^2=4.5$ GeV$^2$.
The ratio of the real to imaginary part, the $\rho$
parameter, is presented in the Fig.~(\ref{ampsplot}-b). The analysis was
done considering the light quarks (the solid lines) and also including the
charm content (the long-dashed lines). The latter gives a lower overall
normalization for the amplitudes, as referred before. The procedure in
order to obtain the real part of amplitude and the $\rho$ parameter was
taken from Ref. \cite{Frankfurt:2000ez}, where a two-power adjustment to the
imaginary part of the amplitude is made, 
\begin{eqnarray} {\cal I}m\,
{\cal A}\,(s,t=0) = A_1\,\left( \frac{W^2}{W_0^2} \right)^{\lambda_1} +
A_2\,\left( \frac{W^2}{W_0^2} \right)^{\lambda_2}\,, \end{eqnarray} 
and the real part and $\rho$ are given by, 
\begin{eqnarray} {\cal R}e\, {\cal
A}\,(s,t=0) & = & A_1\,\left( \frac{W^2}{W_0^2} \right)^{\lambda_1} \,
\tan \left( \frac{\pi\,\lambda_1}{2} \right) + A_2\,\left(
\frac{W^2}{W_0^2} \right)^{\lambda_2}\tan \left( \frac{\pi\,\lambda_1}{2}
\right)\,,\\ \rho & = & \frac{\tan (\pi\,\lambda_1/2) + A_2/A_1\,
(W^2/W_0^2)^{\lambda_2-\lambda_1}\,\tan(\pi\,\lambda_2/2)}{1+ A_2/A_1 \,
(W^2/W_0^2)^{\lambda_2-\lambda_1}}\,.  \end{eqnarray}

The parameters obtained from a 3-flavor (4-flavor) analysis are presented
in Table (\ref{table1}). The effect in the DVCS total cross section is of
order 12 \% in the overall normalisation at $W=75$ GeV ($\rho \approx 0.35$). 
From Fig.~(\ref{ampsplot}-b), the $\rho$ parameter depends weakly
on $W$ for $W\geq 100$ GeV, tending asymptotically to $\rho\approx 0.3$.

\begin{table}[t] \begin{center} \begin{tabular}{lll} \hline \hline
Parameter & (3-flavor) & (4-flavor)  \\ \hline \hline $A_1$ & -0.056 &
 -0.0336373 \\ \hline $A_2$ & 0.0368 & 0.0285722 \\ \hline $\lambda_1$ &
-0.060 & -0.202454 \\ \hline $\lambda_2$ & 0.1059 & 0.152216 \\ \hline
$W_0^2$ & 0.3746 & 8.28475 \\ \hline \hline \end{tabular}
\caption{Parameters of the two-power adjustment to the imaginary part of the
amplitude, with a 3-flavor (4-flavor) analysis.} \label{table1}
\end{center} \end{table}

% Ratio DIS/DVCS
% ----------

\begin{figure}[t]
\centerline{\psfig{file=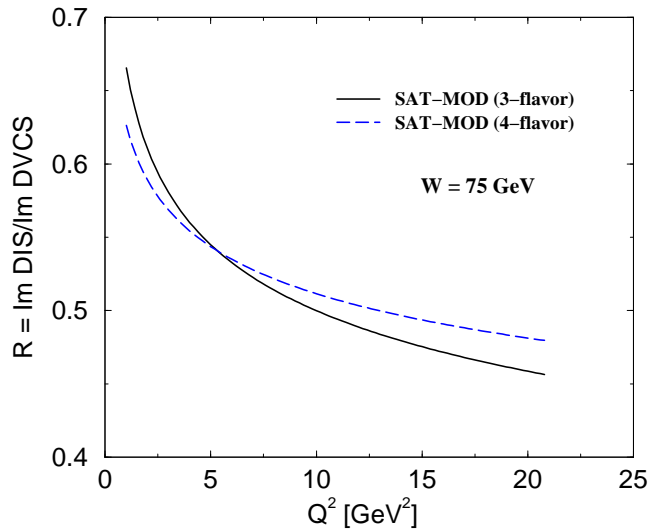,width=85mm}}
\caption{\it The ratio of the imaginary parts of the DIS and DVCS amplitudes
as a function of $Q^2$ at fixed $W=75$ GeV. The solid lines
correspond to the 3-flavor analysis and the long-dashed ones to the 4-flavor
analysis.}
\label{ratiodd} 
\end{figure}

Finaly, we calculate the ratio between the
imaginary parts of the forward $t=0$ amplitudes for DIS and DVCS,
$\mathrm{R}={\cal I}m\, \,\mathrm{DIS}/\,{\cal I}m \,\,\mathrm{DVCS}$. This
quantity is quite interesting as it shows the importance of the skewing
effect. In Fig.~(\ref{ratiodd}) are shown the results obtained from the
3-flavor (solid line) and 4-flavor (dashed line) analysis of the saturation
model at fixed $W=75$ GeV as a function of the virtuality $Q^2$. The
obtained values are slightly above those from an aligned jet model analysis in
Ref. \cite{Frankfurt:1997at} and below those from the dipole analysis in
Ref. \cite{McDermott:2001pt}.

\section{Results and conclusions}

We are able now to compare our results with the H1 Collaboration DVCS 
measurement at the photon level \cite{Adloff:2001cn}. 
The final expression for the
DVCS cross section is written as, \begin{eqnarray} \sigma(\gamma^*
\,p\rightarrow \gamma \,p) & = & \frac{1}{B}\,\left. \frac{d\sigma}{dt}
\right |_{t=0} \,,\\ \left. \frac{d\sigma}{dt} \right |_{t=0} & = &
\frac{[\,{\cal I}m\,{\cal A}(s,0)\,]^2}{16\,\pi} \, (1+\rho^2)\,,
\end{eqnarray} 
where $B$ is the $t$ slope parameter and comes from a simple
exponential parametrisation and the $\rho$
parameter has been determined in the previous section. 
The $B$ value has never been measured for DVCS. The H1 publication 
compares its cross section measurement to several predictions 
for $5<B<9$ GeV$^{-2}$ \cite{Adloff:2001cn}. As a first calculation, we present the
results considering the saturation model for the 3-flavor analysis. 
%1.6 This choice is more preferable since the measured virtuality 
%1.6 $Q^2=4.5$ GeV$^2$ stays below the charm threshold. 
The plots present an upper (lower) value corresponding
to the slope values $B=5$ GeV$^{-2}$ ($B=9$ GeV$^{-2}$).  
In Fig.~({\ref{sigmas}-a) the DVCS cross section is shown as a function
of $W$ at fixed $Q^2=4.5$ GeV$^2$ and in Fig.~({\ref{sigmas}-b) as a 
function of $Q^2$ at fixed $W=75$ GeV.
We have
verified that the inclusion of charm produces a normalization slightly lower than using only
the light quarks where one considers the same $B$ value.  This feature is more
pronounced in the cross section as a function of energy, whereas is less 
significant in the $Q^2$ behaviour. 
Concerning the choice of the quark mass values it should be noted that,
taking $m_f=0$ increases the cross sections for the 3-flavor (4-flavor)
analysis by a factor 1.15 (1.17) at $W$=20 GeV down to 1.13 (1.16) at $W$=120
GeV at the average $Q^2=4.5$ GeV$^2$. This factor increases slightly for
lower $Q^2$ values.
Keeping $m_f$ fixed enhances the overall normalization for lower $m_c$,
as usual.

As a second calculation, we present the
results considering the saturation model for the 4-flavor analysis. 
The plots present an upper (lower) value corresponding
to the slope values $B=5$ GeV$^{-2}$ ($B=9$ GeV$^{-2}$).  
In the Fig.~({\ref{sigmasc}-a) the
DVCS cross section is shown as a function
of $W$ at fixed $Q^2=4.5$ GeV$^2$ and in Fig.~({\ref{sigmasc}-b) 
as a function of $Q^2$ at fixed $W=75$ GeV. 
The conclusions are similar as in the 3-flavor case.

\begin{figure}[t] 
\begin{tabular}{cc} 
\psfig{file=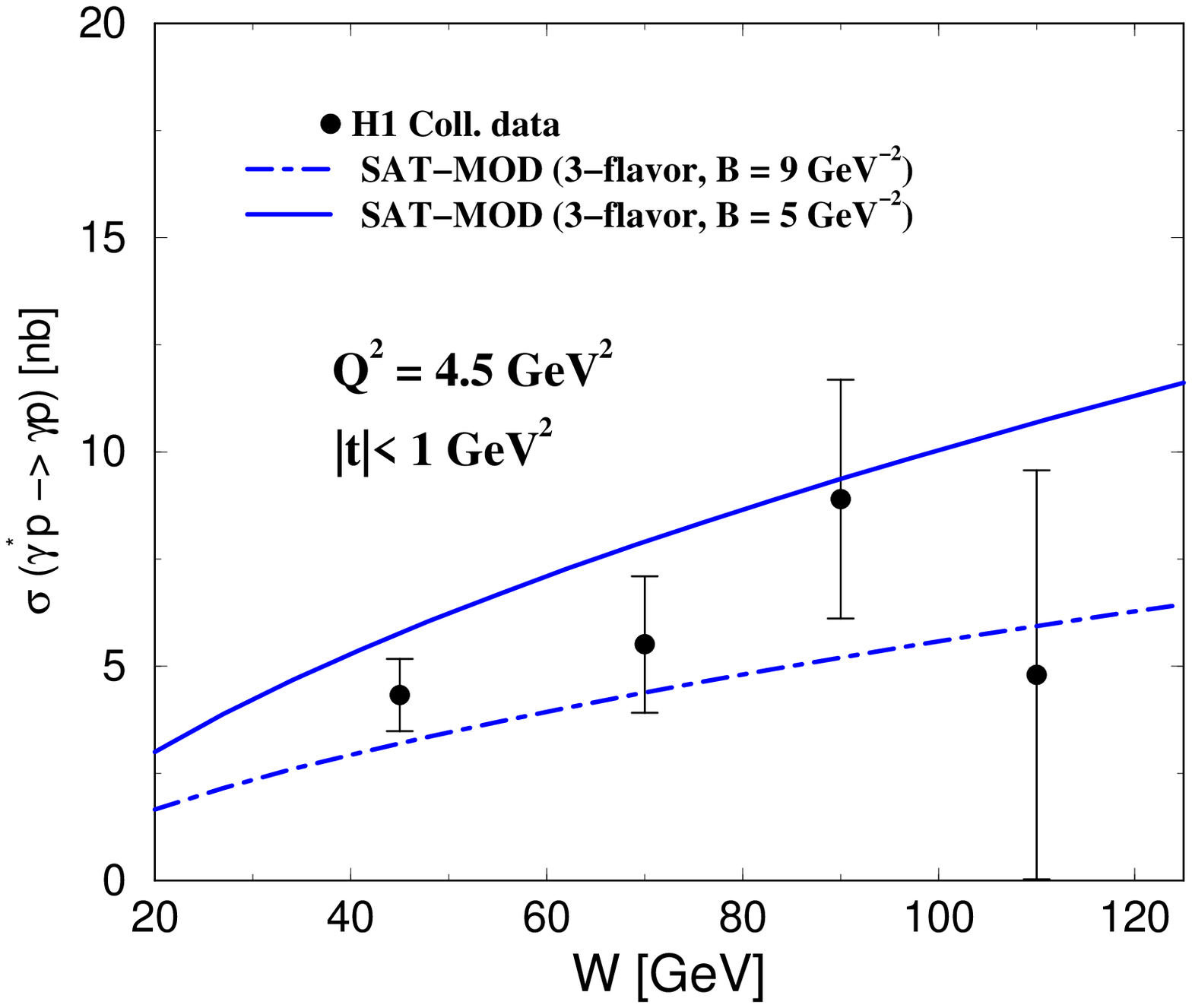,width=80mm} & \psfig{file=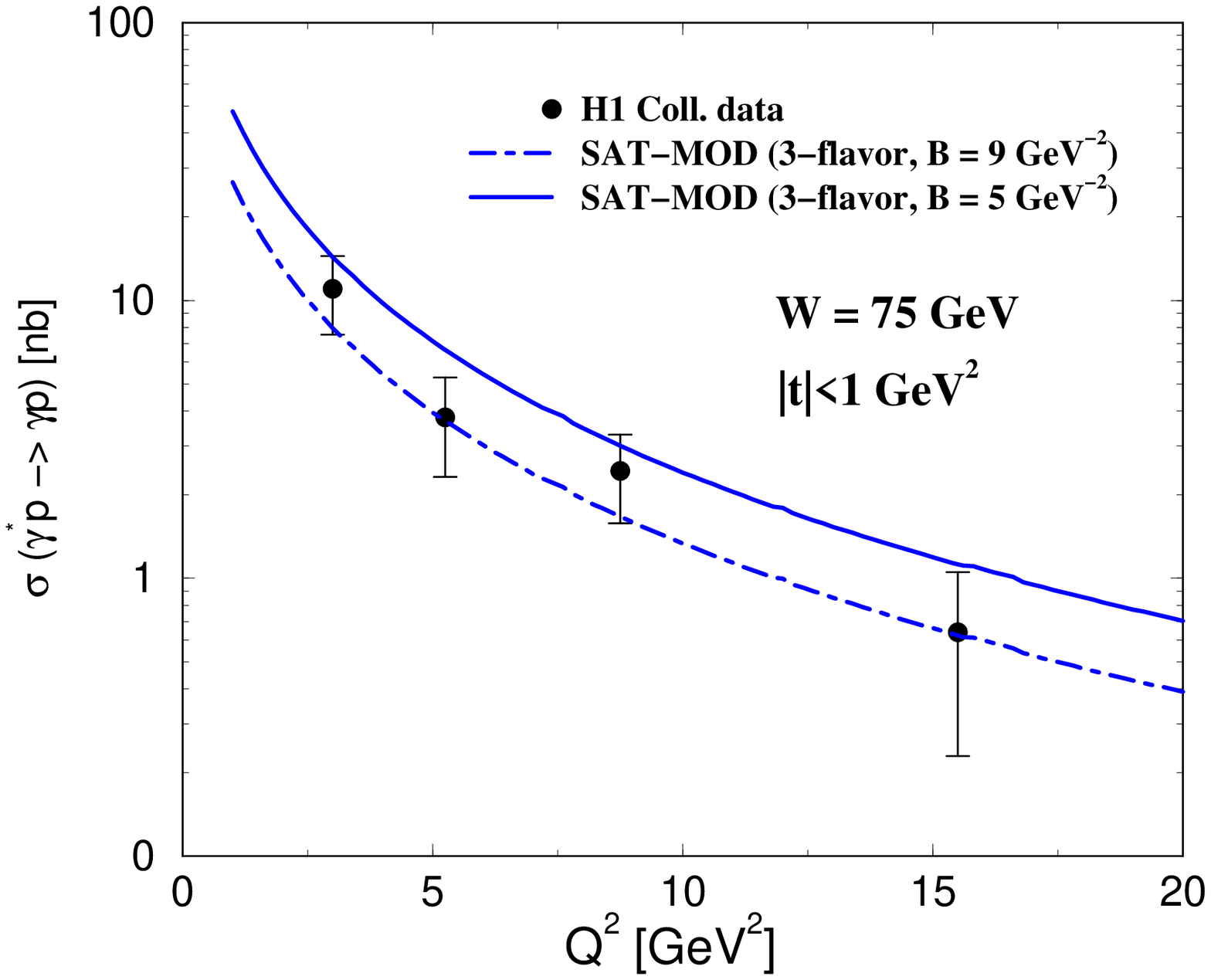,width=80mm} \\ 
\end{tabular} 
\caption{\it {\bf (a)}
Saturation model: a 3-flavor analysis.  The photon level DVCS cross section as a function of the c.m.s. energy
$W$ at $Q^2=4.5$ GeV$^2$ considering the band for the slope $5<B<9$ GeV$^{-2}$ .
{\bf (b)} The DVCS cross section as a function
of the virtuality $Q^2$ at fixed $W=75$ GeV, considering the same notation.  } 
\label{sigmas}
 \end{figure}

\begin{figure}[t] 
\begin{tabular}{cc} 
\psfig{file=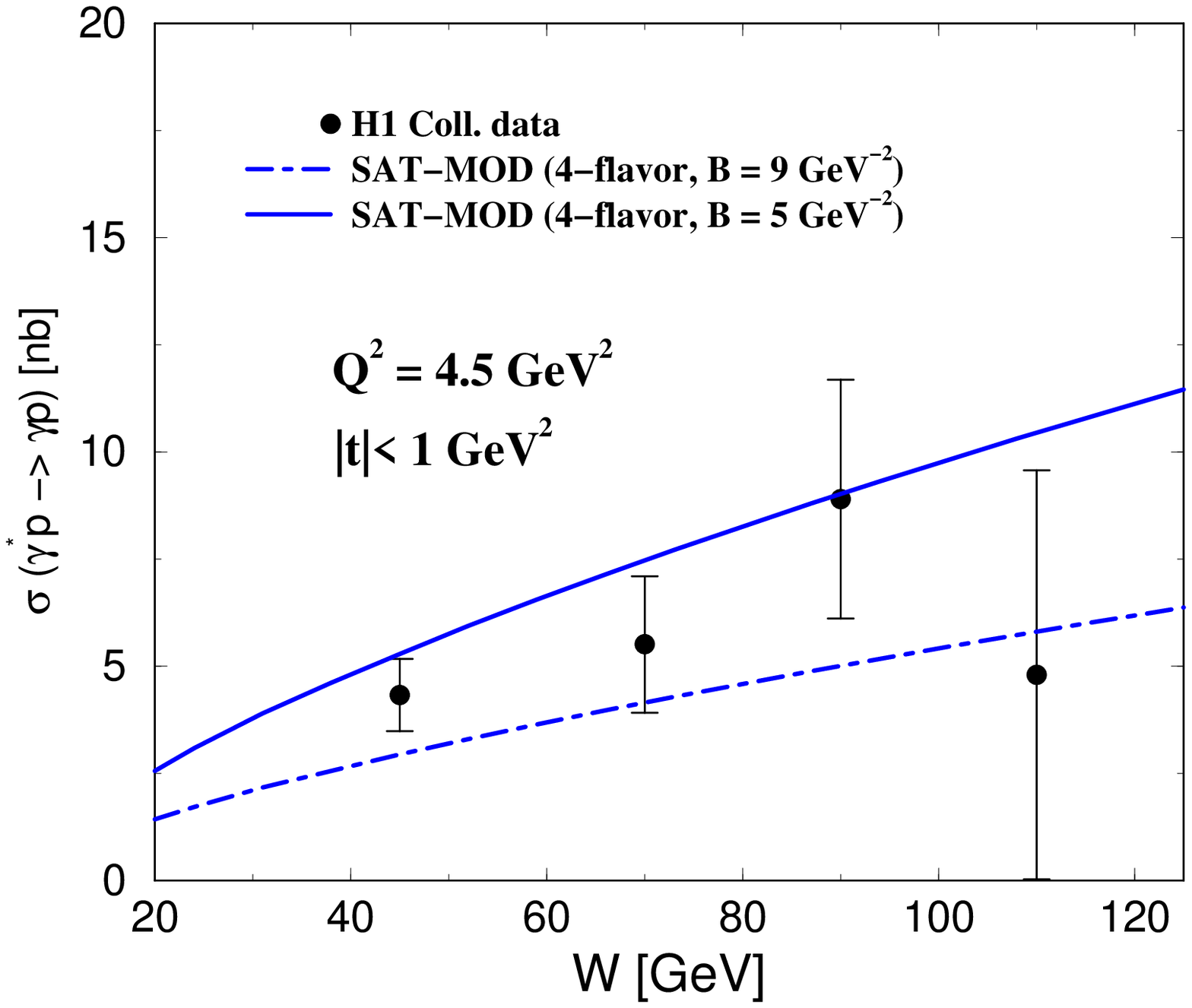,width=80mm} & \psfig{file=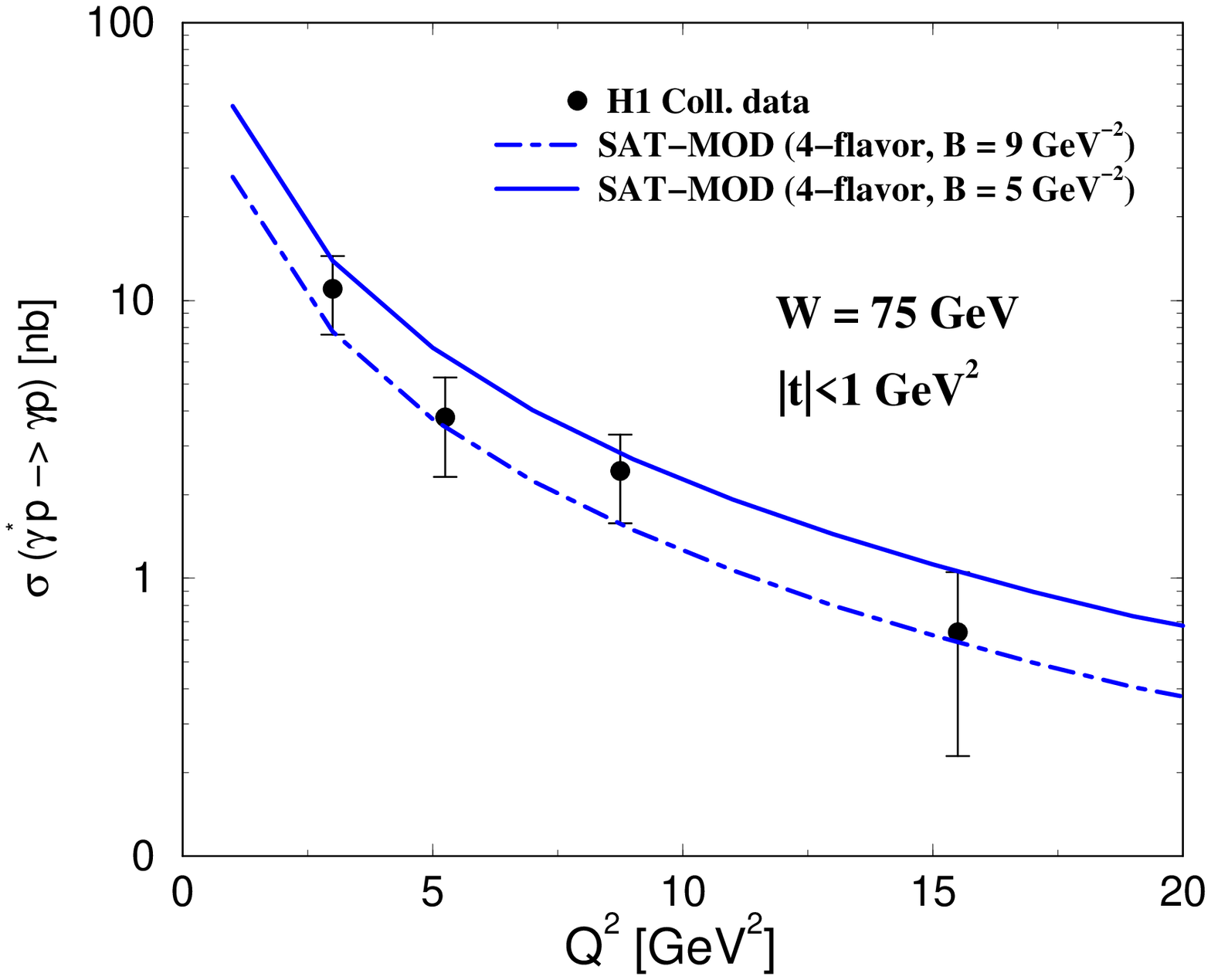,width=80mm} \\ 
\end{tabular} 
\caption{\it {\bf (a)}
Saturation model: a 4-flavor analysis. The photon level DVCS cross section as a function of the c.m.s. energy
$W$ at $Q^2=4.5$ GeV$^2$ considering the band for the slope $5<B<9$ GeV$^{-2}$ .
{\bf (b)} The DVCS cross section as a function
of the virtuality $Q^2$ at fixed $W=75$ GeV, using the same notation. } 
\label{sigmasc}
 \end{figure}

In what follows, due to the absence of a consistent constraint on the
normalization, we will use the $B$ value which fits the best with the H1
measurement for each prediction. In Fig.~(\ref{compmodels}) 
the following calculations are presented: (i) the saturation model for a
3-flavor analysis (solid line); (ii) the saturation model for a 4-flavor
analysis (dashed line) ; (iii) the QCD improved saturation model
(dot-dashed line), given by Eq. (\ref{bgkdip}). The corresponding slope
values are shown in the plot. We have not included the real part of
the amplitude for the BGK case. The present result shows how important it is
to measure the $B$ slope in order to discriminate among the available
approaches. It should be stressed that the result from the BGK model 
deserves more studies, since a 4-flavor analysis was not done yet and 
skewed effects needed in DVCS are not available in this model.

\begin{figure}[t] 
\centerline{\psfig{file=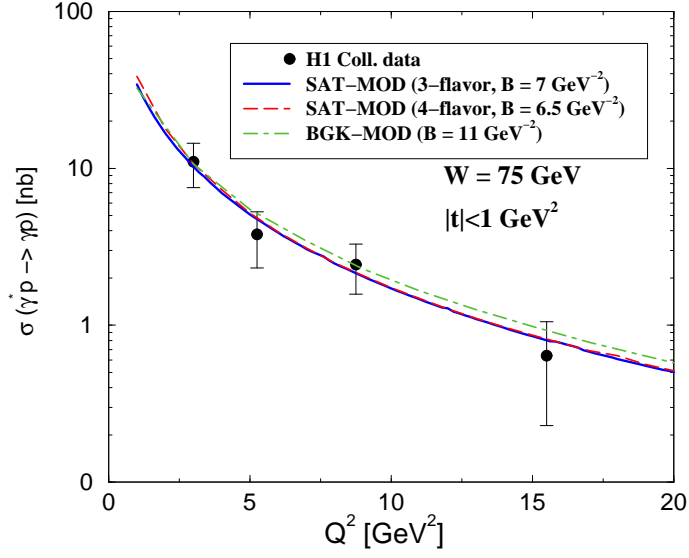,width=90mm}}
\caption{\it {\bf (a)}
The photon level DVCS cross section as a function of virtuality $Q^2$ at
fixed $W=75$ GeV. The plots correspond to the 3-flavor and 4-flavor
analysis of the saturation model (solid and dashed lines) and the BGK
model (dot-dashed line).  The corresponding $B$ slopes are also
presented.}
\label{compmodels}
\end{figure}

% Qualitative analysis
% ----------
It is timely to perform a qualitative analysis in order to obtain a
clear physical interpretation using the saturation model, as done for 
the DIS case in~\cite{Golec-Biernat:1998js}. In the
following we should consider for simplicity the particular case
$m_f=0$. The imaginary part of the DVCS amplitude is given by the
integrations over the dipole size $r$ and longitudinal momentum fraction
$z$ in Eq. (\ref{dvcsdip}).  One can use the small
approximation argument for the Bessel function $K_1(x)$, such that
$K_1(x)=1/x$ for $x\ll 1$ and is exponentially suppressed for $x\gg 1$. 
Then, we can write:
\begin{eqnarray}
{\cal I}m\, {\cal A}\,(s,t=0) \sim \int_0^{4/Q^2} \frac{d\rrn^2}{\rrn^2}
\, \sigma_{dip}\,(\tilde{x},\,\rr^2) + \int_{4/Q^2}^{\infty} \frac{d\rrn^2}{\rrn^2}
\,\left(\frac{1}{Q^2\,\rrn^2} \right) \sigma_{dip}\,(\tilde{x},\,\rr^2)
\label{qualy1}
\end{eqnarray}
where the first term is domainated by symmetric contributions,
%1.6 where the first term corresponds to the symmetric contributions,
i.e. the quark and the antiquark constituting the dipole share almost the same
momentum fraction $z$ and the distribution on that variable is uniform
around the mean value $z=1/2$. The second term 
is domainated by aligned jet or asymmetric
%1.6 corresponds to the aligned jet or asymmetric
configurations, where $z$ or $(1-z)$ takes a vanishing value and the
$z$ integration is restricted to $z \lsim 1/(Q^2\,\rrn^2)$.
%1.6 $z$ integration is restricted to $z<1/(Q^2\,\rrn^2)$.

In order to proceed in the qualitative analysis, one can use an 
approximate form for the dipole cross section from the saturation
model; $\sigma_{dip}=\sigma_0\,\rrn^2\, Q_s^2/4$ for
$\rrn^2 \leq 4/Q_s^2$ and $\sigma_{dip}=\sigma_0$ for $\rrn^2>
4/Q_s^2$. The saturation scale is given by $Q_s^2(x)=1/R_0^2 \simeq
x^{-\lambda}$. Therefore, if the typical dipole size, $1/Q$, is much smaller
than the mean distance between partons $R_0^2(x)$, the expression for
the amplitude becomes,
\begin{eqnarray}
{\cal I}m\, {\cal A}\,(s,t=0)&  \sim &  \int_0^{4/Q^2} \frac{d\rrn^2}{\rrn^2}
\, \left( \frac{\sigma_0\,\rrn^2\,Q_s^2(x)}{4} \right) +
\int_{4/Q^2}^{4/Q_s^2(x)} \frac{d\rrn^2}{\rrn^2}
\,\left(\frac{1}{Q^2\,\rrn^2}  \right)\, \left(
\frac{\sigma_0\,\rrn^2\,Q_s^2(x)}{4} \right) \nonumber \\
& & + \int_{4/Q_s^2(x)}^{\infty} \frac{d\rrn^2}{\rrn^2}
\,\left(\frac{1}{Q^2\,\rrn^2}  \right)\,\sigma_0\,,
\label{qualy2}
\end{eqnarray}
where the first term comes from symmetric configurations and the
remaining ones from the asymmetric configurations. The final result in
this case reads as,
\begin{eqnarray}
{\cal I}m\, {\cal A}\,(s,t=0) & \sim &  \frac{\sigma_0\,Q_s^2(x)}{Q^2}\,\left[ 1+
\log \left( \frac{Q^2}{Q_s^2(x)} \right)\,\right] \,,\label{qualy3}\\
\sigma(\gamma^* \,p\rightarrow \gamma \,p) & \sim &
\left(\frac{Q_s^2(x)}{Q^2}\right)^2 \simeq 
\frac{x^{-\lambda}}{Q^4}= \frac{W^{4\,\lambda}}{Q^4}\,,
\end{eqnarray}
where the large logarithm in Eq. (\ref{qualy3}) comes from the
intermediate region $2/Q < \rrn < 2 R_0(x)$ and the remaining comes from
the regions $\rrn<2/Q$ and $\rrn > 2R_0(x)$. The last expression gives
the DVCS cross section at large $Q^2$ and small $x$.

The other interesting case is when the typical dipole size, $1/Q$,  is larger
than the mean distance between partons, $R_0(x)$. 
The contributions to the amplitude now become,
\begin{eqnarray}
{\cal I}m\, {\cal A}\,(s,t=0) \sim \int_0^{4/Q^2} \frac{d\rrn^2}{\rrn^2}
\, \left( \frac{\sigma_0\,\rrn^2\,Q_s^2(x)}{4} \right) +
\int_{4/Q_s^2(x)}^{4/Q^2} \frac{d\rrn^2}{\rrn^2}
\,\sigma_0 + \int_{4/Q^2}^{\infty} \frac{d\rrn^2}{\rrn^2}
\,\left(\frac{1}{Q^2\,\rrn^2}  \right)\,\sigma_0\,,
\label{qualy4}
\end{eqnarray}
where the two first terms correspond to symmetric configurations and
the last one to the aligned jet configuration.  Performing the
integrals over the dipole size, the amplitude can be written as:
\begin{eqnarray}
{\cal I}m\, {\cal A}\,(s,t=0) & \sim &  \sigma_0\,\left[ 1+
\log \left( \frac{Q_s^2(x)}{Q^2} \right)\,\right] \,,\label{qualy5}\\
\sigma(\gamma^* \,p\rightarrow \gamma \,p) & \sim & \sigma_0^2\,\log^2
\left( \frac{Q_s^2(x)}{Q^2} \right)  \simeq \, {\cal C}\, \log^2 \,(W^2)\,,
\end{eqnarray}
where now the logarithmic contribution comes from the intermediate
region $2R_0(x)<r<2/Q$ and the remaining from regions $r<2R_0(x)$ and
$r>2/Q$. The last expression above gives the behavior of the DVCS
cross section at low $Q^2$. The qualitative results above can be contrasted with the DIS case, where  the $\gamma^{*} p$ total cross section grows as $\sim W^{2\lambda}/Q^2$ at high $Q^2$, whereas at low $Q^2$ it behaves as $\sim \ln\,(W^2)$.  It is worth mentioning the scaling of the
amplitude in the ratio of the variables $\tau=Q_s^2(x)/Q^2$, which
is called the geometric scaling in the DIS case.  
%1.6 provides the geometric scaling in the DIS case.   

In conclusion, our results are comparable with the color dipole analysis 
in~\cite{McDermott:2001pt}, obtaining a good agreement with DVCS cross
sections on the $W$ and $Q^2$ dependences. We have tested the
saturation model in the exclusive DVCS process, showing its
consistency.  The utilized dipole cross section interpolates
successfully the soft and hard regions, producing a smooth transition
between them. In particular, the sensibility to the inclusion of charm
content seems to be more pronounced compared with the inclusive DIS
case. This can be investigated in more details when measurements with higher 
virtuality will become available. A simple qualitative picture for the
DVCS process can be drawn, presenting a clear scaling property.  The QCD
improvement provides good agreement with data requiring a higher $B$
slope in contrast with the original model. The present analysis also
shows the importance of a measurement of the $B$ slope for the DVCS
process.

\section*{Acknowledgments}
The authors thank the support of the High Energy Physics
Group at Institute of Physics, GFPAE/IF-UFRGS, Porto Alegre. 
The authors are grateful to Markus Diehl for helpfull discussions and comments. 
The work of L. Favart is supported by the FNRS of Belgium (convention
IISN 4.4502.01).

\end{document}